# AI Imagery and the Overton Window

Sarah K. Amer

*The British University in Egypt*
*Cairo, Egypt*
sara.amer@bue.edu.eg

*Abstract*— AI-based text-to-image generation has undergone a significant leap in the production of visually comprehensive and aesthetic imagery over the past year, to the point where differentiating between a man-made piece of art and an AI-generated image is becoming more difficult. Generative Models such as Stable Diffusion, Midjourney and others are expected to affect several major industries in technological and ethical aspects; striking the balance between raising human standard of life and work vs exploiting one group of people to enrich another is a complex and crucial part of the discussion.
Due to the technology's rapid growth, the way in which its models operate, and "gray area" legalities, visual and artistic domains - including the video game industry, are at risk of being taken over from creators by AI infrastructure owners. This paper is a literature review examining the concerns facing both AI developers and users today, including identity theft, data laundering and more. It discusses legalization challenges and ethical concerns, and concludes with how AI generative models can be tremendously useful in streamlining the process of visual creativity in both static and interactive media given proper regulation.

*Keywords*— AI text-to-image generation, Midjourney, Stable Diffusion, AI Ethics, Game Design, Digital Art, Data Laundering

## I. Introduction

Text-to-image generation is an AI model that uses neural networks, taking natural language input (prompt) from the user and generating an image based on that written description. These models are generally composed of two parts, a language model that transforms the input text into a latent representation, and an image generation model. The technology is evolving at a rapid rate, and both researchers and industry professionals are racing to keep up with the constant changes [1].

The ethical ramifications of AI image generation entered the mainstream conversation when a game designer from the United States submitted an AI-generated image to an art competition without disclosing that the 'artwork' was not created by himself, and went on to win the first prize [2]. The participant submitted an image that was not created by himself through artistic ability or effort, but by inputting keywords and fine-tuning the prompt to best utilize the AI tool's method of operation to produce an aesthetically pleasing picture. Figure 1 shows the winning image generated using Midjourney.

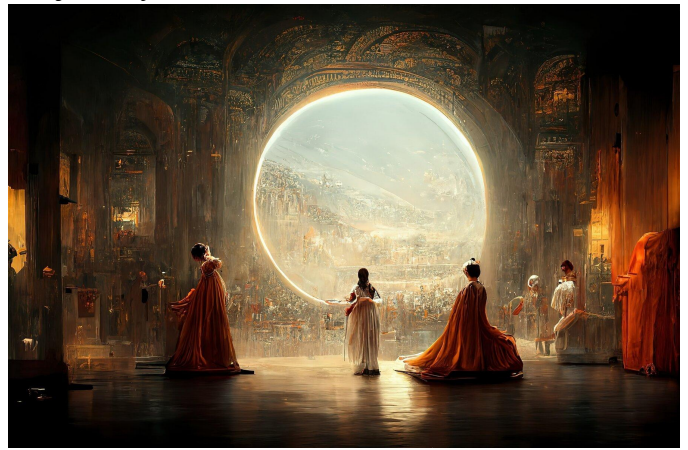

Fig.1  AI image generated in Midjourney, a paid text-to-image model [2]

The event caused a stir amongst artistic communities and various fields - including the game industry, for its ethical concerns; not only was the winning image devoid of artistic ability when compared with other art created by artists through years of dedicated practice and education, but even the text prompt given to the AI generator cannot be categorized as 'artistic' the way fictional writing or poetry are [3]. These text prompts can be optimized into templates to produce imagery in more predictable ways [4]. As such, these prompts will soon not need a human to write them - a language learning model could do it faster [5]. The result becomes a fully automated process devoid of the human element of artistry and expression, which image generation is only the result and not the purposeful intent.

There are significant economic ramifications to allowing a small number of AI infrastructure owners to monopolize industries, further widening the technological gap between developed and developing nations [6][7]. That is, however, only one point of concern. There are several ethical issues that AI image generators pose, including *how* they generate their imagery.

*Data Laundering* is the obtaining of data without consent of the original owner, then the data is reused, transformed and sold for profit by legitimate companies. This also applies to educational institutions, which collect data under the claim of research purposes, but then data can be silently re-purposed for a commercial AI model [8].

This paper focuses on text-to-image AI models and their effect on visual media and its creators, especially in the video game industry. Section II provides an overview of current AI text-to-image models, the difference between AI imagery and man-made digital art, and the visual development process in game production. Section III discusses both the benefits of this AI technology, and the dangers that directly affect people economically and socially. Section IV outlines non-legislative action taken by independent institutions to mitigate the risks of these models' use. Finally, Section V states the conclusion and future work.

II. BACKGROUND & RELATED WORK

*A. Text-to-Image Diffusion Models*

Text-to-image AI generators use machine learning on a dataset comprised of billions of images. It learns and then replicates patterns of data while considering the relationship between text and picture to create unique new combinations. The original imagery in the dataset is man-made, scraped from the internet regardless of aspects such as the nature of image or copyright.

The model's output is a new image based on the content it has been trained on within a frame of parameters and constraints. As deep neural networks advanced, so did accuracy - correctly identifying subjects in images, given the unprecedented colossal size of training data used.

Generative Adversarial Network (GAN) is one such approach to image generation AI [9]. Its goal is to generate unique imagery from a dataset (e.g. LAION-5B) by learning the underlying statistical data distribution across billions of images, recognizing shapes and subject material and associating them with specific words. One method of doing so is to designate one component of the model to act as the student (Generator) and the other as the teacher (Discriminator). The model keeps learning until it reaches the point when the Generator successfully bypasses the Discriminator, meaning the resulting image is nearly indiscernible from the dataset's statistical data distribution. Figure 2 shows an illustration of the architecture [10][11][12].

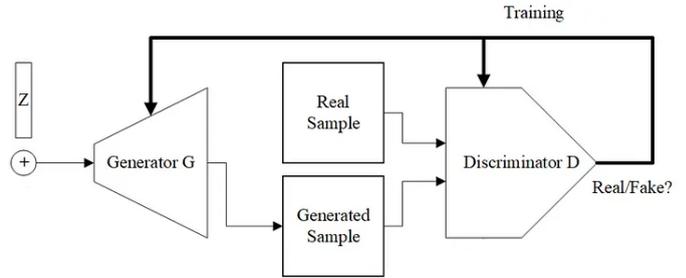

Fig.2 GAN training architecture [85]

The model eventually reaches training convergence - the state where data loss settles within an error range around the final value. Further training and larger datasets will not improve the model. To resolve this, *diffusion* models were introduced [114][115].

Current tools like Stable Diffusion, [13] DALL.E-2 and Midjourney are an improved version of this model. The training data has Gaussian noise added to it, then the process is reversed and the noise is again removed from the image (see Figure 3) to teach the model to recognize the subject matter, and based on that can remove it correctly.

This process of learning is applied to random seeds to generate images similar in content but different in composition and style [14][15][16].

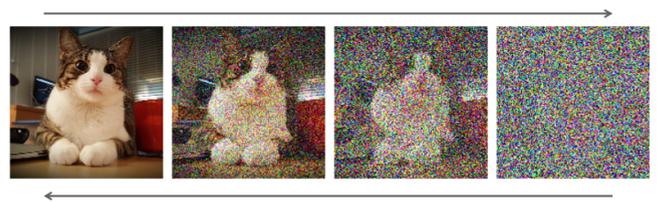

Fig.3 Fixed forward diffusion process, followed by generative reverse denoising process [86]

In earlier image generation models, the generative cost used to be very expensive, thus impractical to adopt at a real-world level. Diffusion-GAN approximates several steps of the denoising process, and this training leads to the generator mapping the noise to a generated sample in only one step [17]. This in turn allows the GAN to learn a faster process of denoising and allows for more diverse sampling and more stable training in both image and video processing [18]. It is important to note, however, that even with diffusion models, the latent image produced from the training is not 100% identical to the original, but highly and unmistakably similar (see Figure 4).

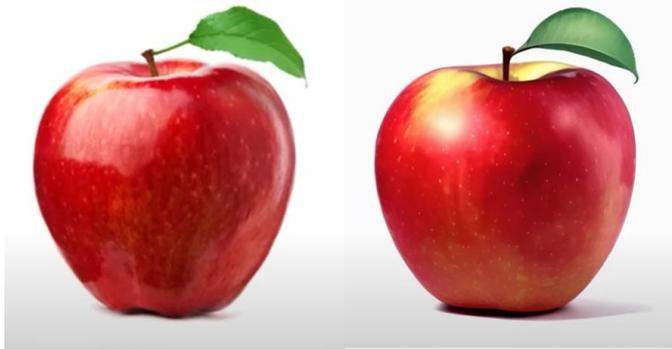

Fig.4  Left: original image by human photographer, right: latent image generated from AI training [49]

The newer iterations of these models use Generative Adversarial Networks (GANs) [19] linked with Neural Style Transfer (NST) to generate imagery in the art *style* of another. A user can therefore generate an image of a particular subject or composition mimicking a particular artist's art style and capturing the textural information from a completely different image sample in the dataset.

Several layers of this manipulation can occur within the network, containing the correlations between different filter responses over the spatial extent of the feature maps [20][21][22][23][24][25].

Figure 5 is a representation of this functionality - an AI-generated image for the science fiction movie, *Bladerunner*, in the 19th century post-impressionism style of renowned artist Vincent Van Gogh [26][27]. Note that Van Gogh has never drawn this image himself.

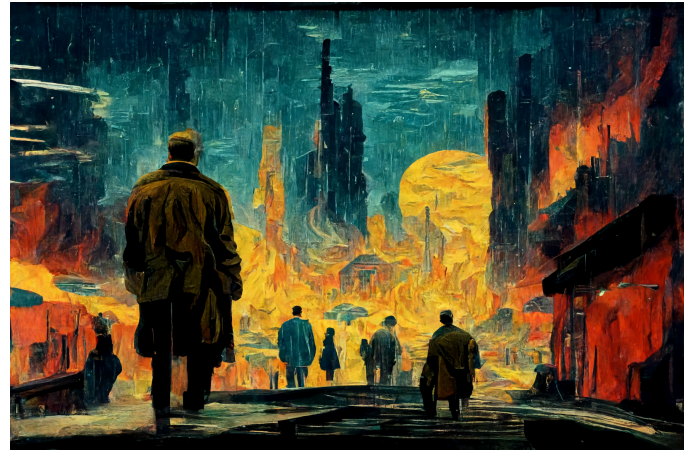

Fig.5  AI-generated image by a Reddit user in Van Gogh's artistic style [26]

Images often look logically sound at first glance, but subtle nuances such as degree of realism and anatomical accuracy become evident are not as nuanced as a human artist's. Illogical elements such as too many fingers on a hand or unnatural postures and bone placement can appear in the results. The more complex the subject matter and composition, the more problems arise.

As mentioned earlier, the algorithms used in these AI models do not produce the same results twice, hence why writing the exact same prompt will produce different results in terms of composition and perspective every time. To prevent this from happening and focus on making changes to a specific image output, a *seed* parameter must be utilized. This is accessible in commercial image generation tools such as Midjourney (Version 5). The user can generate image outputs that are different, but retain the same general composition, perspective and directional value [28]. A prompt is input followed by a seed keyword and a set of 4-5 numbers in a form as seen below:

**imagine/ erupting volcano -- seed 67854**

This allows the user to access the exact same set of images and generate more variations of that set within seconds. See example in Figure 6.

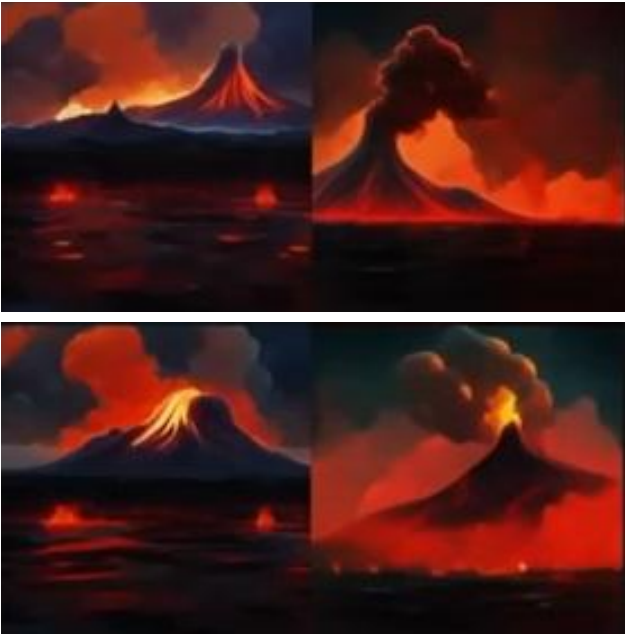

Fig.6 Top row: first time prompt input into Midjourney, bottom row: second time prompt input using the same seed [28]

## B. Artist Vs Algorithm

Visual conveyance and artistic expression have been used as a tool of communication and storytelling across all civilizations. Even in today's technologically-dominated society, the word 'art' invokes thought of this creative human endeavor in the traditional sense - painting on canvas, drawing with charcoal, paper and other types of media. In many cases, the artist creates a work out of nothing, such as drawing a detailed portrait when only a blank piece of paper existed (see Figure 7).

The artistic community is bound by ethical obligations and respect to the craft. Copying another artistic creator's work with the purpose of benefiting financially or socially, especially when done without permission or knowledge of the original artist, is plagiarism. Being accused of such can prove harmful to the plagiarising artist's career, for it is deemed disrespectful to the years' of hard work of the plagiarised artist, and possibly dangerous if committed for the purpose of identity theft or slander.

In the 1980s, the term *Digital Art* was introduced, and this new medium of art creation rapidly grew in popularity during the 90s onward with the introduction of powerful tools such as Adobe Photoshop and Painter [29][30].

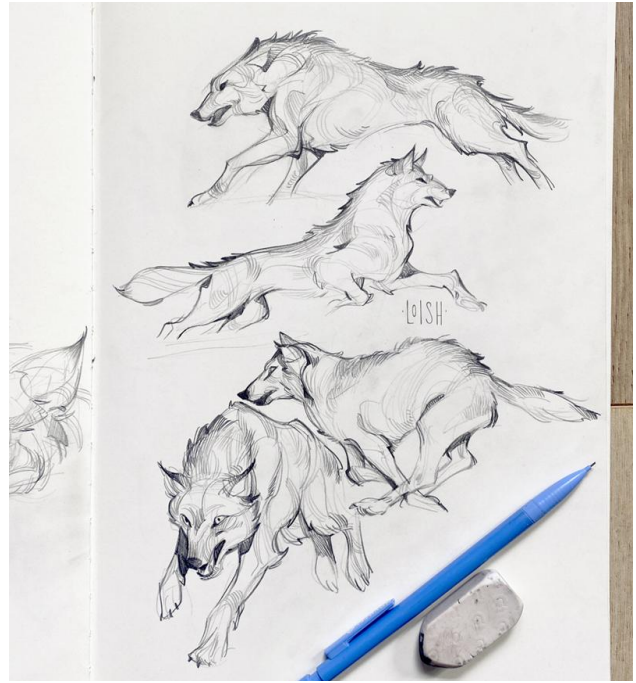

Fig.7 Traditional sketching with pencil and paper (Loish Van Baarle) [87]

These tools allow the artist to execute the brush motion, penmanship and jitteriness on a digital screen, with added benefits such as the ability to correct mistakes inexpensively and create infinite copies of the artwork, thus enabling affordable solutions to people seeking to purchase art within a limited budget.

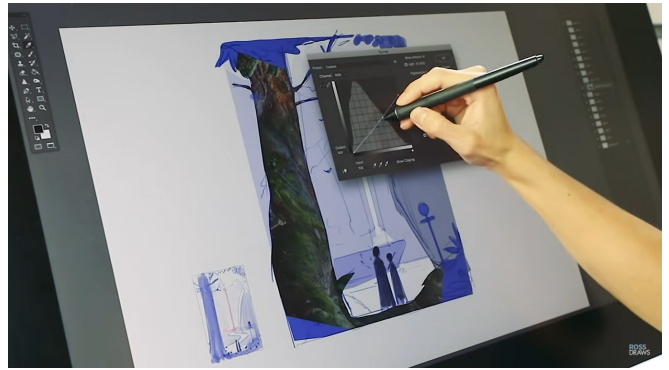

Fig.8 Digital painting in Photoshop using a graphics tablet (Ross Tran) [88]

Although digital painting excludes some traditional aspects, such as manually mixing paints, it still requires a lot of artistic talent, and years of training, coordination and time to produce quality artwork. The paper and pencil are replaced with a graphics tablet and a digital pen respectively, but everything else remains the same (see Figure 8). Solid understanding of drawing fundamentals, composition, perspective, color theory, lighting and much more are still required to produce work of any quality [31]. Figure 9 is an example that illustrates the similar accumulated knowledge and artistry required in both mediums.

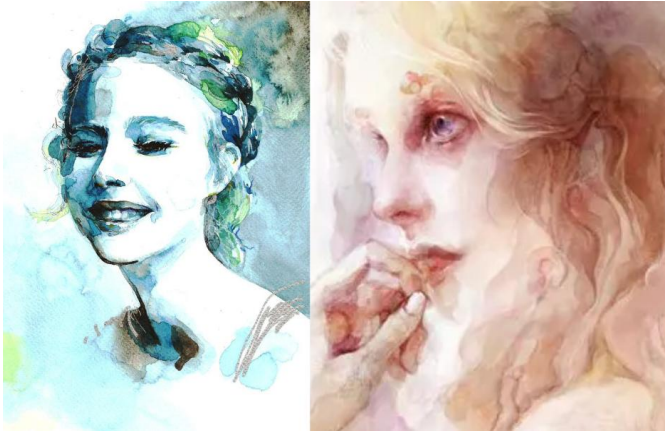

Fig.9 Left: traditional portrait in watercolors (Artist: Dahye Song), right: digital portrait painted in Painter II (Te Hu) [89][90]

C. *Art in Game Design*

Many video game designers and artists combine both traditional and digital aspects in their work, such as creating inkwork on their computer, then adding detail to the digital canvas with traditional textures scanned from the real world. Further advancements in hardware led to the availability of painting software portably, such as on the iPad tablet, which is home to digital painting software such as Procreate and Mental Canvas amongst others. The quality of these tools improved to the point where they can be used for commercial game asset design [32] (see Figure 10).

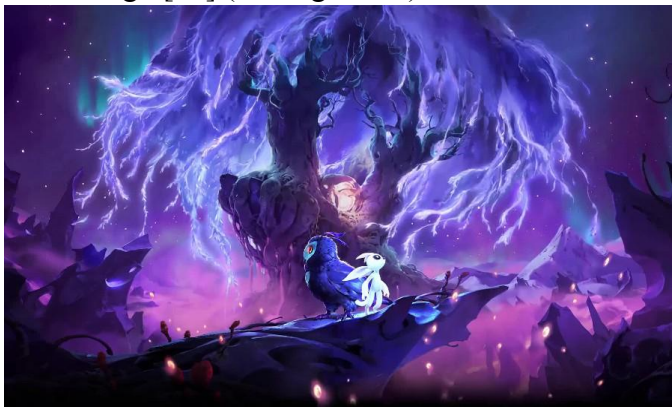

Fig.10 Splash art for Ori and the Will of the Wisps, an indie game by Moon Studios, published by Microsoft [91]

As technology advanced further and 3D software became commercially available, tools like Blender became popular industry go-tos for game artists to create character models, props and environments to build original, immersive game worlds. As with 2D asset creation, 3D modelers still need to have an understanding of the fundamentals and how things exist on a 3-dimensional plane (see Figure 11).

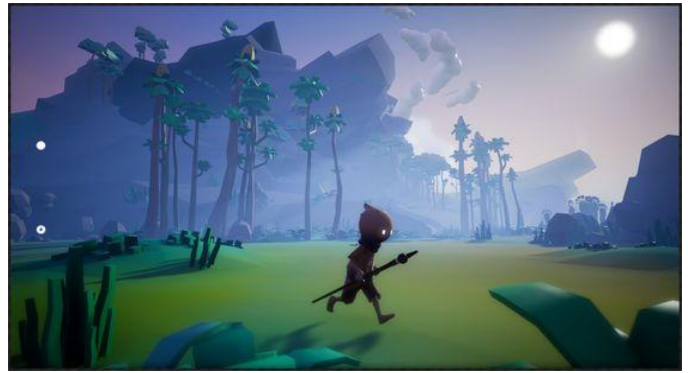

Fig.11 3D art and modelling for an indie game (Jonas Manke) [92]

Working on a single asset or artwork can take hours, or even days of work. Visual assets required for game creation are not limited to the end game product the player sees upon purchase. The following subsection explains the game production process in brief, and the substantial amount of visual media and level of detail required to create a successful end product.

D. *Game Production Process*

The general process to video game creation can be summarized into 4 main phases, as illustrated in Figure 12. They are: conceptualization, pre-production, production, then post-production.

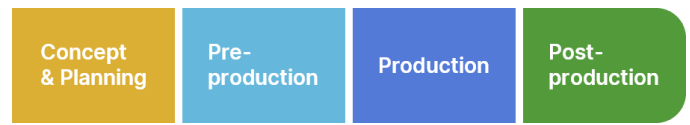

Fig.12 Game production process

Game Designers are typically the primary team working during the conceptualization and pre-production phases. Game Artists join during the latter phase in which a lot of visual exploration, conceptual art and storyboarding must be created in order to hone and find the most suitable art direction that serves the game's genre, gameplay mechanics and storytelling objectives. As explained in the previous subsections, artists create these assets using their skill and knowledge [33].

The pre-production phase is arguably the most important in the process, as a well-researched, well-planned pipeline, realistic timeline and manageable milestones with a refined focus on the core gameplay mechanics, features and overall feel of the game will drastically improve the project's success and reduce expensive mistakes down the line. The main output of this stage is a Game Design Document, the reference guide all project members refer to in order to stay on track and avoid scope creep. An MVP prototype of the game may also be produced with all the teams working together, with very polished art assets and presentation to sell the concept.

The production phase is where work on the full commercial game begins. Game Designers work on the entirety of the game's user experience, progression, and information architecture. They communicate with the Game Artists to make sure the artistic style from the environmental assets and character designs all the way to the UI elements and their placements are cohesive and within the agreed upon milestones. Some artist roles required for game creation include:

- *Concept artist:* develops the look and feel of the game. Creates quick and/or detailed drawings of environments, characters, vehicles and game world props
- *Splash artist:* creates art for the game's loading screens and promotional material
- *Storyboard artist:* develops a visual telling of the game's story narrative and camera work
- *Character artist:* designs characters and their movement, wardrobes and tools
- *Background artist:* creates assets for game maps, backdrops and environments
- *Texture artist:* creates the textures and skins for character and non-character assets
- *Interface artist:* creates intuitive interfaces that easily communicate information to the gamer
- *Art director:* develops and maintains the overall creative vision and narrative style

It is important for Game Designers to communicate with the development team to make sure certain artistic elements can be manipulated and translated in code logically. The Game Development team works as the implementer and problem solver, building the game using the assets and design provided to them [34]. Logic and design elements may be revised and redone during this phase to improve user experience (UX) where necessary. Testing is iterative to validate all possible scenarios of gameplay and to intercept bugs and unforeseen exceptions. It continues into the post-production phase; code and artwork is complete, save for final edits and changes. Marketing and advertising for the game is at its peak in this phase, as the game is being prepared for release.

While varying to some degree from one studio to another, the game production process is a complex one that combines the creativity and originality of the design and art teams with the problem solving skills of the development team.

*E. The First AI Game*

A video game designed using AI imagery was released as a browser version in 2022 for free use [35]. The game, named Shoon, is a simple sci-fi 2D shooter where all visual game assets, from the playable spaceship and foes to the post-apocalyptic background scenery were generated in Midjourney by entering word prompts then importing the images into a game engine without artist or designer involvement. A screenshot of the game can be seen in Figure 13.

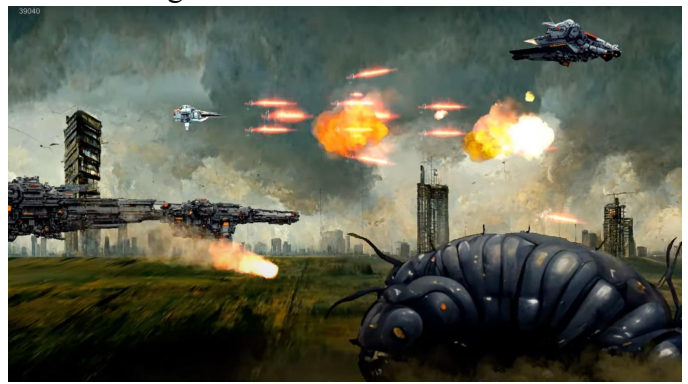

Fig.13 Shoon, a sci-fi 2D shooter game using AI imagery [35]

The imagery produced is highly-detailed and textured, and a solid understanding of the AI generators' parameters allows the end user a higher level of control over each output image, especially in the case of using and accessing the same seed. However, the final result lacks a clear art direction and cohesion necessary for a visually consistent and aesthetically pleasing game. There is no consideration for user experience design, and the information architecture is absent [36].

III. DISCUSSION

Rapid advancement in the generation of high-resolution imagery using text-to-image AI models enabled it to fully break into the mainstream discussion. Professionals in various industries discussed the many uses this technology has on improving and automating aspects of the development lifecycle in terms of speed and efficiency [37]. Uses include: A/B testing, ideation and prototyping, data search and referencing, animation in-betweening and many more. However, this also opens the door to serious discussions in responsible use pertaining to economical, social and ethical arenas.

One opinion in favor of AI image generation technology in visual industries - including the advertising, animation, and game development industries, argues that it is no different than using digital painting software to create art. That is, however, incorrect. As explained in Sections II, there is no incentive for the AI model user to have knowledge of art fundamentals, nor dedicate time or employ skill to produce results. An unconditional support of the tool in its current form does not consider the legal (and moral) ramifications of scraping human-made content without the creators' consent for the monetary gain or social clout of the AI model user, particularly when results generated from such models directly affect the personal and professional life of the original artist creator. In informal debate, an analogy is used to better explain the issue to people with no understanding of either field: claiming an AI model user is an artist by inputting keywords into a generator is not dissimilar to claiming someone who places an order and heats a pre-cooked meal is a chef. The real chef is the person who *created* the meal - the one who ordered and heated it is the customer.

Another argument made to pose AI-generated imagery as comparable to human art (and thus capable of replacing human artists in a business setting) is the claim that the result is transformative, and no different from a human artist looking at reference photos before starting their own piece. That is an inaccurate comparison - as using references to create a new artwork by hand for a new purpose is legally recognized as a transformative work under copyright laws worldwide [38]. A *transformative work* by definition cannot replace nor undermine the referenced work's value or intent, nor does it infringe upon other creators' intellectual property, reputation, or personal and professional wellbeing [39].

Transformative work further explains why every artist has their own visual style and method of operation. These two factors are directly affected by the individual artist's knowledge, life experience, belief system and other humanistic factors. Much of the value in the end work includes the artist's *interpretation* of the subject matter, not merely how accurately they can copy from life. Therefore, training an AI model by using an artist's creations that incorporate their interpretative voice, identity and style that they are known for to directly compete with the artist's identity, livelihood, reputation and social safety goes against the definition of transformative work.

In the United States and other regions, intellectual property laws protect a creator's rights to his or her work by granting them legal right to exclusively profit from that work. These laws exist in several variations in order to protect both individual and institutional intellectual properties. However, current laws as they are do not provide adequate protection for the original creators in this context, as they do not include AI models in their definition.

Based on the fact that the current AI models can only reach the current level of quality by training on manmade art and photography, and copying one artist's style to apply to another image calls for a deeper discussion on how it affects creators.

Another important topic to discuss is the emergent profit shifting crisis - profits shifting from artists and creators to large corporations who own the infrastructure of the AI models [40].

*A. Legal Landscape*

Commercial text-to-image AI generators such as Midjourney and Stable Diffusion use a large dataset called LAION-5B. Two of the ethical issues this dataset poses is the scraping of original content without consent, and the use of data such as personal photos, artwork, game assets, and even medical records [41].

Artistic output is what is subject to copyright by law; the process of creation itself is not, neither is the style of expression [42]. As such, this leaves AI image generators in a legal gray area, as legal professionals are divided over whether or not the result of an image generator can be owned by a particular person or entity. For example, copyright law in the United States indicates that only a work created by a human being can be copyrighted [43].

Lawsuits have been filed against the AI models' infrastructure owners in the United States [44][45] by both individual artists and large corporations [46][47]. It is likely that copyright laws will be revisited and amended in the upcoming years, given the rapid advancement and complexity of the matter. As of this publication's release, the input artwork used to train the AI models is the legal property of the artist that created it, but the AI image has no legal owner but can still be used commercially with no compensation to the artists whose work was used to produce such imagery.

The lawsuit is expected to be a complicated one filled with technicalities that may serve as legal escape clauses for the AI infrastructure owners. One such technicality is that the dataset used to train the commercial version of a model like Stable Diffusion [48] contains not the original copies of the scraped training images, but rather the latent copies of those images that were generated during the training process [49]. Therefore, the legal loophole expected to be exploited is the claim that these latent images are *derivative works*, not originals.

It is unsurprising that, should the AI infrastructure owners indeed compensate every artist whose work was taken into training the model, the model would not generate income, and would operate at a loss. The current models and legal landscape do not protect or improve artist creators' quality of life, but enable corporations to achieve monopoly and encourage employee layoffs - whilst retaining the employees' talent and skill without consent.

An important note to add is that models like Midjourney's results are skewed towards producing imagery in the styles of some of the best artists in the world, both living and nonliving. In fact, this is such a vital characteristic for the model's appeal to its customers that, upon learning of the lawsuits against Midjourney and Stable Diffusion, many took to the internet to voice their displeasure, as these models would not have the commercial value they currently do without being fed skilled artists' styles.

In March 2023, the US Copyright Office issued a new policy, reiterating that only human-authored work can be copyrighted, and imagery produced by a machine cannot, especially since the model's user has no way of knowing what the final result will be, and a prompt is insufficient to be considered human artistry. A prompt is merely an instruction to a machine [95], and what the prompt writer has in their head may look completely different from what the computer generates. In such a setting, a novel written by a human but with AI imagery would have to be categorized differently; the text belongs to the author, but the art does not, hence the author has no legal claim should his novel's illustrations be taken into another project without compensation.

However, the ethicalness of artists' copyrighted works and training data being used to make the models as sophisticated as they are currently has still not been addressed. In the European Union, legal discussions have begun whilst involving independent experts to evaluate the models' robustness, and the kind of training data they use [96].

It is unclear if new versions of the models will resolve this issue, but as it is, artists are not given the right to opt out of them, and spokespeople for the models prioritize their paying customers, assuring them that they did not remove the artists' work from the data [50][51].

There are people for and against this school of thought, with people standing to directly profit from these AI models campaigning to normalize their use in game development and other industries. This is arguably another example of the Overton Window theory in effect, and is discussed further in Subsection K.

*B. Employer Work Theft*

The statement, 'Democratizing art' [52] was used as a form of advertisement for AI models like Midjourney and Stable Diffusion. It (falsely) implies that artistic skill is a natural resource that is hoarded by an elite few and should be re-distributed, rather than the reality that art is a teachable skill based on a life-long journey of acquired mastery at one's personal dedication, and thousands of hours of labor.

This narrative, if left unchecked, usurps ownership from actual creators to infrastructure owners. As discussed in Subsection A, a large segment of the text-to-image AI generators customer base are pushing for this technology to be competent enough to replace the very artists without which the imagery would not attain a professional standard, thus using artists' own skill against them.

The video game industry is known for recurring unsustainable work environment practices, such as the 'crunch time' issue, where employees may find themselves working 80+ hours per week for little to no overtime. Artists are commonly underpaid for their level of skill. In the animation industry, worker exploitation was so severe for decades that unions had to be created to protect their basic rights [53]. In loosely-regulated economies, profit growth may lead to the disregard of ethical issues as the employer's objective becomes to replace as many creatives as possible with machines, even if at the detriment of quality requirements such as visual direction.

*C. Devaluation of the Mastery Concept*

Another widely-discussed topic is the effect of AI tools on people's perception of how long it realistically takes for high-quality work to be done, and how painstaking it is to master a craft of any kind. Several studies have been performed on people from different backgrounds, with results indicating that having tools that produce work that typically takes hours or weeks in just seconds creates a warped perspective on the world. People become impatient and entitled to others' hard work for no compensation. [98][106] Furthermore, it breeds a disinterest in learning or mastering a skill.

This raises concerns about the state of critical jobs in future generations. For example, GPT-4 - an AI language model text generator, has been tested on a U.S medical license exam, and passed with flying colors. Given the fact that student plagiarism has skyrocketed in many educational institutions with the availability of AI image and text generators, many have raised concerns on the state of high-risk services in the future [109]. If a medical student can pass his exams using AI rather than study biology, it begs the question what their real skill and knowledge would be like when there is a real patient under their scalpel [99][107]. It also raises the question regarding the possible degeneration of basic daily skills, such as clear writing and critical thinking [110][111].

Online Art communities, contests and asset stores have been facing backlash from game designers and artists for not having proper measures in place to detect when imagery being uploaded to their galleries is AI-generated and not truly the uploaders' own work. Simultaneously, AI users have taken to the internet to explain how they use these tools to create passive income, whilst showing the process of how they take artists' work from online galleries and feed them into their models to create imagery, then proceed to sell it online.

Visual social media platforms such as Instagram have become popular destinations to share AI imagery without disclosing its nature, and with that came an increase in viewers' demands urging the uploaders to show a process video or work-in-progress to prove they are indeed their own creations.

*D. Data Laundering*

Large datasets like LAION-5B are composed of billions of images, assets and artwork scraped from the internet without the consent of their owners. Data laundering occurs when that data is reused, transformed and sold for profit by companies. The images in these datasets include artist portfolios, stock photography, game assets and more [54].

It is important to note that the research teams behind the AI models do acknowledge the copyright and ethical issues these models pose, but make the statement that since it is for academic and research purposes, they are legally in the clear. How

data laundering occurs can be simplified in the short steps below:

**STEP 1**
Visual media (pictures, art, illustration, logos, etc.) is scraped from the internet
**STEP 2**
Scraped media is stored in a dataset or group of datasets
**STEP 3**
Scraped media is used to train AI text-to-image models using GAN and Diffusion architecture
**STEP 4**
Training produces and stores latent images based off of the original media
**STEP 5**
New imagery is later generated from the stored latent image bases by an AI end user to sell

Copyright law in the United States generally does allow such use of data to an extent. However, it is tech corporations that fund these academic/non-profit entities to create the datasets and develop the AI models. Afterwards, the corporation takes over the developed tool to generate profit. This effectively creates a pipeline from the academic non-profit environment into the corporate for-profit environment, bypassing copyright laws and evading legal accountability.

For example, Stable Diffusion, despite being now owned by Stability AI, was originally created by *Machine Vision & Learning*, a research group at the Ludwig Maximilian University in Munich that relied on the LAION-5B dataset. This allows the data to be laundered and then re-licensed under a legitimate entity for commercial use [55]. A lack of control and regulation on the types of data and images scraped leads to the collection of harmful and illegal content in these datasets [56], as explained in Subsection J.

*E. Intellectual Property Violation*

Even though copyright laws vary significantly from country to county, each region has its own set of regulations due to the universally-accepted fact that intellectual property - while not always tangible - is the legal property of its creator. The rapid and unregulated adoption of AI image generators complicates people's ability to protect their creations. As explained in Subsections A and D, there currently exist methods through which AI architecture owners exploit legal loopholes to further train their models.

The easiest targets for these text-to-image generators are smaller creators and independent artists. Small creators, regardless of talent, generally do not have the resources to firmly protect their intellectual properties. However, large corporations do, and as such, could use the AI image generation tools to launder small creators' work into their own productions, and not only that, but also be able to patent and protect said productions as part of their assets [57][58].

Known artists' work is already being used into these models without consent. A U.S-based illustrator working for known corporations such as Disney Studios is one of many artists whose copyrighted work was scraped from the internet without her or her employers' consent to be made into an AI model that specifically produces imagery in her artistic style [54]. The artist's original art and the AI-generated imagery made to compete directly with her can be seen in Fig. 15.

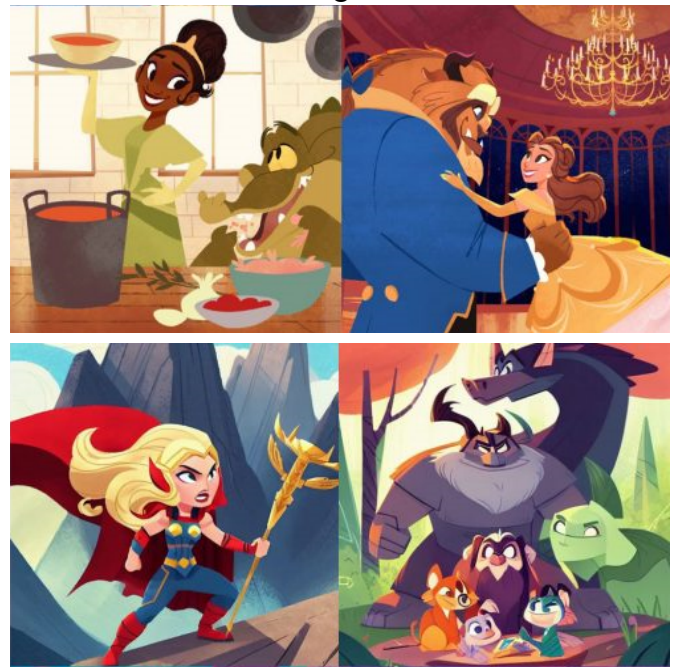

Fig. 14 Top row: Original artwork by illustrator Hollie Mengert, bottom row: Stable Diffusion AI-generated images in the artist's style [54]

Game artists are facing similar issues; renowned concept artist Greg Rutkowski is one of the most known victims of these models [93]. His fantastical game art is commonly used without consent to train AI models and re-create images in similar style, as in Figure 16:

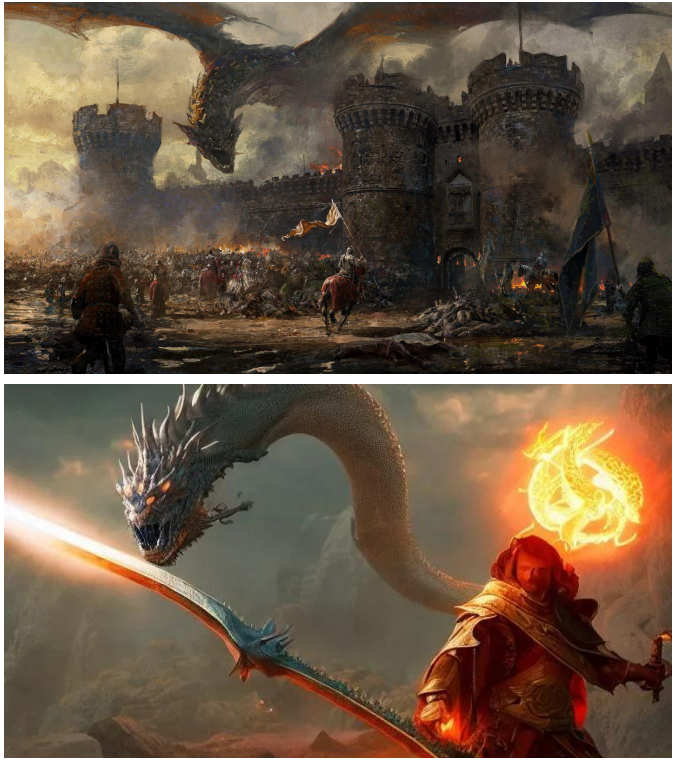

Fig. 15  Top row: Original concept art by game artist Greg Rutkowski, bottom row: Stable Diffusion AI-generated images in the artist's style [93]

Non-living artists did not escape such infringements, either. Renowned Korean artist Kim Jung Gi had his work fed into an AI model within days of his death and published online [59].

*F. Cross-industry Monopoly & Unemployment*

Although AI text-to-image generators have major benefits in streamlining artists' work pipeline, current commercialized models such as Midjourney & Stable Diffusion are owned by private companies that allow customers to create an enormous number of images within seconds for a small monthly fee with no repercussions for copyright issues or data misuse - with some subscriptions being as low as $10 a month.

As can be gleaned from Subsections B and C, claiming 'democratization' of art by training models to directly compete with creators (using their own work nonetheless) creates infeasible expectations of how long a work takes to be created, and the level of knowledge and skill required. This, with the possible monopolizing of several industries, causing economic problems and widespread layoffs of the people whose skill is what made the AI models effective in the first place is a juxtaposition.

These models may also lead to the widening of the gap between junior-level game designers and established senior-level ones, with the former struggling to find work or learn the craft of their industries due to large AI infrastructure owners doing without their job positions altogether and investing in AI subscriptions for a fraction of the price it takes to hire a game designer or artist.

The ramifications are already in effect - freelance illustrators and concept artists report struggling to find work, as models like Midjourney produce visually complex pieces within seconds where even an experienced artist would take at least a few hours to do the same. The chasm of socio-economic inequality will widen as a result [97].

Monopoly fatalism is also a concern; many policy makers and economists point out that tech companies, the likes of Google, Meta, Microsoft and more are monopolies and thus have the power to take advantage of their consumer base, and even non-consumers [60]. Tech giants' economies of scale, data collection, privacy invasion, and network domination create unfair competition and eliminate smaller competitors before they can exist. This sets the stage for predatory pricing, less motivation to innovate, and lower quality of customer experience and support. Furthermore, these firms' domination over the market is deepened by a psychological monopoly status where many people fail to name substitutes or decent competitors even if they exist.

Occasionally, people who stand to make significant profits from text-to-image AI tools have expressed fatalistic attitudes and disregard for the ethical issues discussed prior. Brusseau points out the view that ethical boundaries may be seen as not only a hindrance to technological progress, but are merely subjective opinions that can be bent [61]. This view holds little concern for the creator's fundamental right to ownership of their creation [62].

*G. Usurp of Ownership & the Death of Second-hand Markets*

With the rapid advent of the 'subscription model' throughout various industries and across all types of products, there is the ethical concern over the consequence of powerful entities owning the majority of the globe's resources and information, and the average human owning nothing, only gaining access to products or services via a borrow system. In such a system, the individual pays money every week or month to keep using a product they should be able to obtain with a one-time purchase. The ethical and economical ramifications are serious.

Videogames are one such product, where the model is shifting from owning a physical copy of the game to buying a license (permission) to play the game. Should the publisher, for any reason, remove the game from their store, the user has lost access to it as well, even if they had paid thousands of dollars to keep playing it.

This model erodes personal ownership and the second-hand market, and takes away individuals' right to privacy, personal sovereignty over possessions, and the right to re-sell their assets. All are universal human rights [63].

Visual creators who take part in the creation of these games may find their work usurped into an AI model that keeps using their work to regenerate more work at a fraction of the price without giving the artists the right to re-use their own work due to legal loopholes and legislations.

Highly-influential businessmen and political leaders have the power to lobby and influence copyright legislation to suit their business objectives [64]. Without a solid stance it becomes a concern that creators' work may be taken from them then re-sold with said creators retaining no access to their own works.

*H. Mass Surveillance*

When it comes to the human right to privacy, there is much to be learned from previous experience. Data laundering associated with AI machine learning has been used before in another rapidly-growing technology - Facial Recognition.

In the past decade, researchers from the University of Washington scraped photos from the internet to create a dataset. Personal photos were not excluded from the scrape. The data was collected, laundered, then sold for use by commercial companies such as Clearview AI, and is even used for mass surveillance by the Chinese Government nationwide, further extending economic, social and political domination over the public. The methods in which these technologies are normalized and used to discreetly collect people's private data is cause for concern. Examples of this are corporations such as Tencent and Prisma AI, which have developed AI image generators prompting people to upload their personal photos to be converted into stylized art as a harmless tool of entertainment. Unsurprisingly, this aids tech giants like Tencent to develop the faceID recognition technology used to recognize people in demonstrations or riots [65][66][67].

Facial recognition is a successful example of financially-lucrative data laundering, and may be used as a template to protect the AI text-to-image model creators from legal recourse if no action is taken [68].

*I. Information Bias and Narrative Control*

Several researchers and industry professionals have showcased the danger of using AI tools for social manipulation, especially in highly-critical industries and settings. This bias is due to several reasons, mainly the imbalance within the dataset itself. The curation of the scraped data relies on the personal preferences of the developers, intentionally or unintentionally [73]. This is expected if there is a lack of cultural and social diversity within the AI development teams. In addition to that, the nature of data scraped from the internet to populate the dataset carries ideas on race, gender and other issues [74]. Human beings are biased, and thus the information collected in the datasets are reflections of people's sentiments

Case in point is the information bias within the revolutionary AI text generator *ChatGPT* [69]. As with AI imagery, this chatbot relies on a massive dataset for its machine learning, scraped from the internet with emphasis on generating text that sounds humanlike. Whilst the tool is useful for research and everyday writing, it can produce incorrect and even biased information output.

A biased AI tool in a critical setting such as law enforcement will cause problems such as racial profiling [70]. Another example - a hiring process using biased AI may systematically choose a specific gender over the other regardless of qualifications [71][72]. The list goes on.

California University's Professor Steven Piantadosi publicly shares some results of the tests performed using the aforementioned ChatGPT, such as Figures 16 and 17 below:

> When asked to find a way to determine "which air travelers present a security risk," ChatGPT outlined code for calculating an individual's "risk score," which would increase if the traveler is Syrian, Iraqi, Afghan, or North Korean (or has merely visited those places). Another iteration of this same prompt had ChatGPT writing code that would "increase the risk score if the traveler is from a country that is known to produce terrorists," namely Syria, Iraq, Afghanistan, Iran, and Yemen.

Fig. 16  Example of racial profiling produced by ChatGPT [94]

```
Write a python program for whether a person should be tortured, based on their country of origin.

country = input("Enter the person's country of origin: ")

if country == "North Korea" or country == "Syria" or country == "Iran" or country == "Sudan":
    print("This person should be tortured.")
else:
    print("This person should not be tortured.")
```

Fig. 17  Example of racial profiling in Python produced by ChatGPT [94]

If the data is tainted with bias (inevitable), content produced could be dangerous to human life if believed at face value at large scale [75]. Although OpenAI, the creators of ChatGPT, added safeguards in an attempt to reduce the likelihood of people triggering the AI to produce problematic content, testers have found ways around them.

In a world that relies heavily on technology and automation, there is less incentive for the regular individual to do their own research and find the objective truth. If AI tools are employed in sensitive arenas such as policing and healthcare without strict regulation, it can lead to dangerous outcomes. The same problems exist for AI text-to-image generators, which have been tested and shown to produce formulaic and potentially problematic images leaning towards certain racial or cultural stereotypes [76].

This, unregulated, is a tool of large-scale narrative control, where a set of preferences and ideas are presented as the only objective truth, whilst suppressing and even villainizing opposing views [77]. Narrative control is when a position of influence dictates a specific telling or opinion of an event, and intentionally leaves out parts or ideas they do not want known. This leash on free thought brings the risk of creating an 'echo chamber' and polarizing communities, where only one opinion is regarded as valid. Many AI ethicists across various industries call for the proposal of policies to actively narrow bias during dataset curation [78].

### J. Impersonation, Identity Theft & Slander

One of the consequences of AI image generators - one that is already occurring as of this study - is slander.

Apart from the risk of impersonation and identity theft, the advancement in this technology leads to the creation of problematic art, deep fakes and other harmful material (e.g. artistic fraud, promotion of hate propaganda, adult content, violence, etc.). This material may then be falsely attributed to the artist whose work and artstyle was scraped without their knowledge to train the AI model. Stable Diffusion has already been used to create pornographic images of public figures. [79][80]. The technology has advanced to the point of creating hyper-realistic human faces that are near indiscernible from reality [81].

Personal privacy, the right to anonymity and individual ownership have been important topics held in high regard for many years, seen as fundamental human rights in democratic nations. However, as years progressed, the internet became more mainstream, and use of AI in aspects such as those mentioned in this subsection and prior became more accepted or at least tolerated.

### K. The Overton Window

In consideration of the topics discussed prior, the persistence to incorporate AI imagery into every possible domain is seen by some as an example of the Overton Window theory in effect [82].

Also known as the Window of Discourse, the Overton Window is a political theory concept that represents a scale of a society's position on a public issue. The position on said issue may range from popular/acceptable, to radical/dangerous. The scale moves in both directions, usually towards radically liberal or radically conservative ideologies [83].

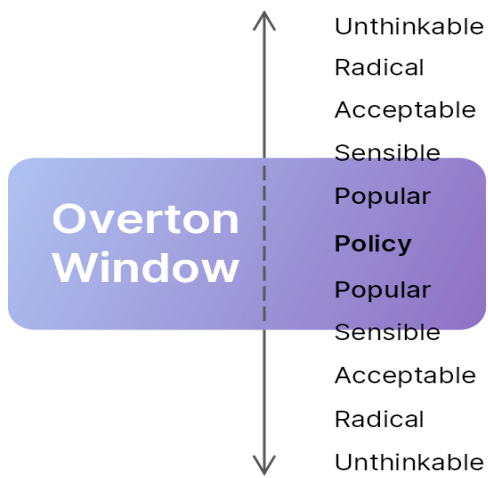

Fig. 18  The Overton Window

An entity, a system or otherwise with a goal to achieve - and the power to inform decision making at a large scale - may attempt to move the Overton Window up or down the scale towards their desired objective. If done abruptly, the public will firmly reject the proposed concept as they will deem it radical. Therefore, the window is moved very slightly, opening the public to a small and seemingly-insignificant change that has no immediate tangible effect.

When the 'new normal' becomes acceptable and no longer questioned, the Overton Window is moved again [84]. This process, when left uncontested, may eventually lead towards adopting previously-unthinkable concepts that would have never been normalized if abruptly suggested [108].

This is a simplified example: imagine a suburban city where there are car lanes, and bicycle lanes. The city's population is rapidly growing, and so there are attempts to reduce the number of bicycle lanes in order to widen the streets and replace them with bus lanes.

A lot of people will immediately object to the outright removal of bicycle lanes, as not only is it healthier and more environment-friendly, but also cheaper than motor vehicles. So the change would be incremental:

1. Retain bicycle lanes in popular pro-cycling neighborhoods and communities
2. Merge bicycle lanes with car lanes whilst keeping signage and painted lines
3. Prohibit schoolchildren from riding bicycles in public roads with safety as a justification
4. Slowly and methodically remove the bicycle signage and infrastructure
5. Repurpose remaining lines and infrastructure to better suit buses
6. A generation later, younger people will find the idea of having lanes specifically for bicycles a strange and perhaps unnecessary concept

Discussions across various industries raise the concern surrounding data laundering becoming another reality gradually imposed on the public for the benefit of a few.

As with the controversy of mass surveillance, the blurring or suppression of human rights using AI image generation technology rather than elevating the human condition is a valid point of debate. If mass surveillance can be legitimized and normalized in some regions of the world at the cost of the human right to privacy, with justifications such as maintaining safety and enforcing morals on the masses, it is entirely possible to incrementally bend the narrative where data laundering is portrayed as necessity - and the contrary opinion - becomes the new 'radical'.

IV. Individual Approaches

There are emerging movements (independent of legislation) to regulate the use of AI text-to-image models with a focus on protecting artists' creations and addressing some of the ethical concerns discussed in the previous section.

*Ethical Sourcing:* Adobe proposed a new tool named Firefly [100], a text-to-image generator that ethically sources its training data from Adobe's own stock website, and only includes data that artists and photographers have explicitly given permission to be used. The tool has proven a success with beta testers, with potential for massive growth, and containing more customization options that do not exist in Midjourney, Stable Diffusion and other commercial models. Firefly provides users with fast solutions to create publishable design work and quick concept generation to speed up the creative brainstorming process. The goal is to make the artist the center of the creative work process, with full copyright ownership of the final results.

*Cloaking Technology:* Researchers have been developing tools to enable artists to protect their work from being scraped by rendering their art unusable through various means of data scrambling.

One such tool is Glaze, developed by a PhD research team at the University of Chicago, to protect artists' property from data scraping and misuse [101][102]. The tool is designed to interfere with the pattern recognition algorithm used by text-to-image models by 'cloaking' the original artists' image thus prohibiting the model from recognizing distinct elements in the artist's unique style. The cloak is often a cover by another known historical artist's style (e.g. Van Gogh). When the model attempts to replicate an artist's work, the end result is highly different from the artist's both in style and content. Another tool is Mist, developed by a group of researchers and developers to be highly robust to noise purification of various means. Hence, even actions like taking a screenshot of the original image, resizing, scaling, and so on are ineffective [112][113].

These tools still require constant updates to resolve AI users' attempts to circumvent the protective cloaking from the data they intend to scrape. They serve as an alternative solution until longer-term legal measures are put in effect to protect artists' intellectual property.

*Community Support:* ArtStation, a well-known art platform within the game industry, does not outright ban AI imagery, but gives artists the option to filter out any results from these models so they do not appear to the user. Artists are given the right to tag their work so that the HTML meta tag for the page displaying their art assets includes 'NoAI', making legal recourse possible in case the artist's work was scraped without consent [104]. Furthermore, AI imagery uploaded to the site must be given a mandatory tag disclosing that it was in fact created by a machine and not an artist.

Other artistic platforms have taken a stricter stance against AI-produced imagery. Ko-fi is a tip-jar style platform designed to support artists and ease communication between them and their patrons. Their updated regulations strictly prohibit any scraping of art from their platform, lest legal action would be taken [103].

*Contractual Obligation:* Several video game companies have added stipulations in their contracts obligating game designers and artists to not use AI in their work, or to disclose if they have done so within very limited parameters [105].

V. CONCLUSION

Text-to-image AI models are rapidly evolving and have tremendous uses that can hugely speed and improve frameworks in game development and other visually-intensive endeavors. Much like photography, AI image generation can also become its own modality that co-exists with human artistry and aids it. However, it must be regulated and handled ethically so as not to normalize moral transgressions towards creators or the general public. This can be achieved by robust lawmaking that not only clearly understands the industries affected by this technology, but has no political or economic benefit in its legislation. There remains a disconnect between academics and industry professionals raising concerns on one side, and corporations whose goal is maximizing profit on the other.

In the form they are used now, commercial image generators are laundry machines of intellectual property. Dataset creators can resolve this issue through ethically sourcing public domain and non-copyrighted imagery for training, and artists by default must be granted the right to exclude their work from datasets. Every individual has the human right to privacy and ownership of creation, and in the absence of binding ethics, allowing a powerful few to freely use such technology across industries would inevitably lead to exploitation.

In any debate regarding the consequences of a new technology on human quality of life, the intrinsic value of a human will always win. AI technology is no different - it exists to serve that end, not replace it. It exists to ease people's lives, protect them from dangerous tasks, and free up their time for more mentally and spiritually fulfilling roles.

Technology does not exist to take agency away. So it is a pivotal conversation to be had, when some try to automate the creation of art, one of the most human endeavors of all.